\documentclass[10pt,conference]{IEEEtran}
\usepackage[T1]{fontenc}
\usepackage{cite}
\usepackage{amsmath}
\usepackage{graphicx}
\usepackage{booktabs,tabularx,array}
\usepackage{xurl}
\usepackage{tikz}
\usetikzlibrary{arrows.meta,positioning,fit,calc,shapes.geometric}
\usepackage[hidelinks]{hyperref}
\hypersetup{pdftitle={Coding Agents in KVM: Practical Containment and User Control with Limited Resources},pdfauthor={},pdfsubject={},pdfkeywords={AI engineering, coding agents, virtualization, containment, sovereignty, deployment}}
\newcommand{\tool}{KVM-Agent}
\newcommand{\code}[1]{\texttt{#1}}
\newcolumntype{Y}{>{\raggedright\arraybackslash}X}
\title{Coding Agents in KVM: Practical Containment and User Control with Limited Resources}
\author{\IEEEauthorblockN{Yutaka Nagashima\\ Institute of Computer Science, The Czech Academy of Sciences}}
\begin{document}
\maketitle

\begin{abstract}
Coding agents often need broad authority for productive development work, but granting that authority can expose users’ data and infrastructure. Off-the-shelf cloud workspaces can help contain agent activity, but reliance on provider-specific platforms can lead to vendor lock-in and reduced user control. Researchers and engineers must therefore balance agent autonomy, enforced containment, and user control within available financial, computational, and human resources. We analyze these trade-offs and present KVM-Agent, an open-source implementation that automates most setup and provisioning steps. It gives agents broad authority inside replaceable Ubuntu VMs while keeping host administration outside those VMs and supporting multiple coding-agent applications, model services, and remote job delegation. The approach retains familiar workflows and user choice at the cost of VM resources, maintenance, and residual risks from 
secrets accessible inside the VMs and permitted network access.
\end{abstract}


\section{Introduction}
In an ideal world, everyone would have access to coding agents that are safe, affordable, open-source, easy to use, and efficient, without losing control over their data and computing infrastructure. 
In practice, these goals involve trade-offs. 
Broad permissions enable development work but increase the potential damage from mistakes or compromise. 
Cloud tools and model services can reduce local operational demands and offer scalability, but introduce external dependencies. 
Familiar development environments support existing tools and workflows, but deploying and maintaining them on user-controlled infrastructure requires computing resources and human effort.

An agent may need to install a compiler, build an unfamiliar repository, or repair dependencies. Coding agents perform such actions through an agent--computer interface~\cite{yang2024swe}, exercising authority over the execution environment. Mistakes, malicious dependencies, and indirect prompt injection~\cite{greshake2023indirect} can redirect that authority. Running these operations in an ordinary workstation account exposes the files and sessions accessible to that account; requiring human approval for every command makes progress depend on repeated user intervention. At the same time, a deployment that demands extensive security administration, a particular commercial service, or dedicated new hardware may be impractical for a small research group with tight resource constraints.

In this paper, we introduce \tool{} to address the following question:
\emph{How can researchers and engineers give coding agents the authority needed for development work while preserving their choice of tools and infrastructure and keeping operational costs manageable?}

\tool{} is an operational wrapper around existing coding-agent applications.
It provisions a graphical Ubuntu virtual machine (VM).
We call the operating system (OS) inside the VM the \emph{guest} and the OS running the VM the \emph{host}.
Agents use an account with passwordless sudo inside the guest.
The user retains host virtualization controls and a dedicated recovery key.
The design also supports a choice of coding-agent applications and large language model (LLM) service providers, remote operation, and optional job delegation to another machine.
Our contributions are an analysis of these trade-offs in deploying coding agents under resource constraints (Section~II) and an open-source reference implementation of the resulting design (Section~III). The implementation automates most setup and provisioning steps, requiring manual input where user-specific choices, credentials, or trust decisions are needed.

\section{Design Goals and Trade-offs}\label{sec:goals}

We analyze three trade-offs that guide the design of \tool{}:
\emph{agent autonomy versus containment},
\emph{provider reliance versus user sovereignty}, and
\emph{functionality versus operational cost}.

\begin{table*}[t]
\caption{Isolation alternatives on Linux; the selected approach is marked. Boundary and agent authority are separate dimensions, not security rankings. Desktop container products may add a VM; Docker Sandboxes is a separate microVM-based product.}
\label{tab:isolation}
\centering
\small
\renewcommand{\arraystretch}{1.12}
\begin{tabularx}{\textwidth}{@{}>{\raggedright\arraybackslash}p{1.85cm}>{\raggedright\arraybackslash}p{2.8cm}>{\raggedright\arraybackslash}p{3.35cm}YY@{}}
\toprule
Choice & Isolation boundary & Agent authority & Practical advantages & Costs and limitations \\
\midrule
Dedicated bare-metal machine & Separate machine; no trusted workload on the agent's OS. & Entire Linux OS, including root and local devices. & Native performance/device access; no guest-to-host escape; hardware reuse. & Whole-machine allocation; external recovery. Supplied secrets and network access remain exposed. \\
Dedicated Linux account & UID/GID and permissions; shared kernel. & Account permissions; no host sudo. & Minimal setup; native tools/filesystem; no guest OS. & Shared kernel/resources; readable files exposed. System changes need trusted administration. \\
Docker containers & Namespaces, seccomp, resource controls; shared kernel. & Configured user, capabilities, mounts; rootful or rootless. Container root runs with reduced capabilities by default. & Container tooling; lightweight process environments. & Privileged daemon/socket access, mounts, and capabilities can expose the host. Rootless still shares its kernel. \\
Podman containers & Similar kernel boundary to Docker. & Configured user, capabilities, mounts; rootful or rootless. & Daemonless CLI; per-user operation; Open Container Initiative (OCI) compatibility. & Rootless modes in both engines add networking, UID-mapping, and resource-control constraints. \\
\textbf{KVM/QEMU VM}\par\textbf{(Our choice)} & Separate kernel/virtual hardware; trusts host and hypervisor. & Guest root/sudo; host administration stays with the user. & Full OS/desktop; existing tools/installers; disposable disk. & Boot, RAM, maintenance, and device/console attack surface; separate egress policy needed. \\
\bottomrule
\end{tabularx}

\end{table*}

\subsection{Autonomy versus containment}\label{sec:goal-autonomy}
We aim to
\textit{grant broad authority within a replaceable project-specific environment while limiting access to host resources that may contain administrative credentials and other data unrelated to the project.}
Passwordless sudo inside the VM permits system-level installation and repair;
we accept that guest compromise exposes the project's code, data, sessions, and supplied credentials.
To limit the blast radius, our intended deployment assigns one or more VMs exclusively to each project, although the implementation does not enforce this convention.
This trade-off suits research projects where source-code confidentiality is not a primary requirement.

Convenient interaction creates additional channels: host–guest clipboard sharing, exported files, remote-editor forwarding, and worker access all carry data or authority across a boundary. These channels can improve convenience while increasing exposure. The design therefore distinguishes restrictions enforced outside the guest from guest-local defaults and explicit user practices. These considerations include network reachability, the placement of credentials, output review, and user control over external actions, as well as the choice of isolation mechanism.

\subsection{Provider reliance versus user sovereignty}\label{sec:goal-sovereignty}
We aim to help users retain \emph{sovereignty}: their practical ability to choose and administer the execution infrastructure, coding-agent application, LLM service, and data destinations.
The open-source virtualization stack can run on user-controlled hardware, and the guest accommodates both proprietary applications and open-source alternatives. Open-weight inference is an option, not a requirement or a guarantee of equivalent capability.

Using a hosted LLM service can reduce local compute and maintenance requirements, but sends prompts and context to the provider and may incur service charges. Running a model locally places compute and maintenance demands on the user; limited hardware may constrain model choice and thereby affect task quality. Supporting several applications preserves choice but can increase maintenance effort, particularly when their software dependencies or configuration requirements differ.

\subsection{Functionality versus operational cost}\label{sec:goal-operation}
We aim to keep setup and daily operation manageable while supporting standard development tools, remote access, and worker delegation. These functions impose different costs: a full Ubuntu guest requires computing resources and software maintenance, while delegation adds machines and access permissions to manage. The design accepts these costs to accommodate project-specific software and use computing capacity available across existing machines.

To limit administrative effort, repetitive setup and provisioning steps should be automated, and procedures for operation and recovery should be explicit. Manual input is needed for user-specific configuration, authentication, and trust decisions.

\subsection{Comparing deployment approaches}\label{sec:choices}
Table~\ref{tab:isolation} compares alternatives against the preceding goals.
Dedicated bare metal removes the guest-to-host escape path and avoids virtualization overhead, but recovery must be performed outside the compromised OS. It remains a credible choice when spare hardware is available, although physical separation does not restrict network actions or protect supplied credentials.

A dedicated Linux account runs agents under a separate user identity, with file access governed by Unix permissions~\cite{linuxcredentials}. It shares the host kernel; withholding sudo limits autonomous system-level installation and repair, while granting unrestricted sudo defeats account-based separation.

Containers add namespaces and resource controls while sharing the host kernel~\cite{nistcontainers,dockersecurity}. Docker and Podman both support rootless operation; Podman's normal CLI is daemonless~\cite{dockerrootless,podman}.

We select KVM/QEMU primarily to address autonomy versus containment (Section~\ref{sec:goal-autonomy}). 
KVM~\cite{kivity2007kvm} provides hardware-assisted virtualization in the Linux kernel, QEMU~\cite{bellard2005qemu} supplies the virtual hardware, and libvirt~\cite{libvirtoverview} manages VM configuration, lifecycle, storage, and networking.
Agents can administer a full development OS inside the guest, while users retain host administration, recovery, and resource controls outside it. 

Compared with dedicating a physical machine to each project, separate VMs let users share hardware across projects and switch between project environments.
The trusted host retains control over resource allocation and replacement of compromised VMs. 
Host-managed networking can keep physical-network credentials, such as Wi-Fi passwords, outside the VMs. These benefits come at the cost of virtualization overhead and maintenance of the host software stack.

The open-source virtualization stack also supports user sovereignty (Section~\ref{sec:goal-sovereignty}) by keeping the execution environment independent of a particular coding-agent application or model provider. For functionality versus operational cost (Section~\ref{sec:goal-operation}), the full guest OS accommodates existing development tools, while our setup automation and recovery procedures aim to keep deployment and administration manageable on available hardware.

\section{Architecture and Reference Implementation}\label{sec:system}
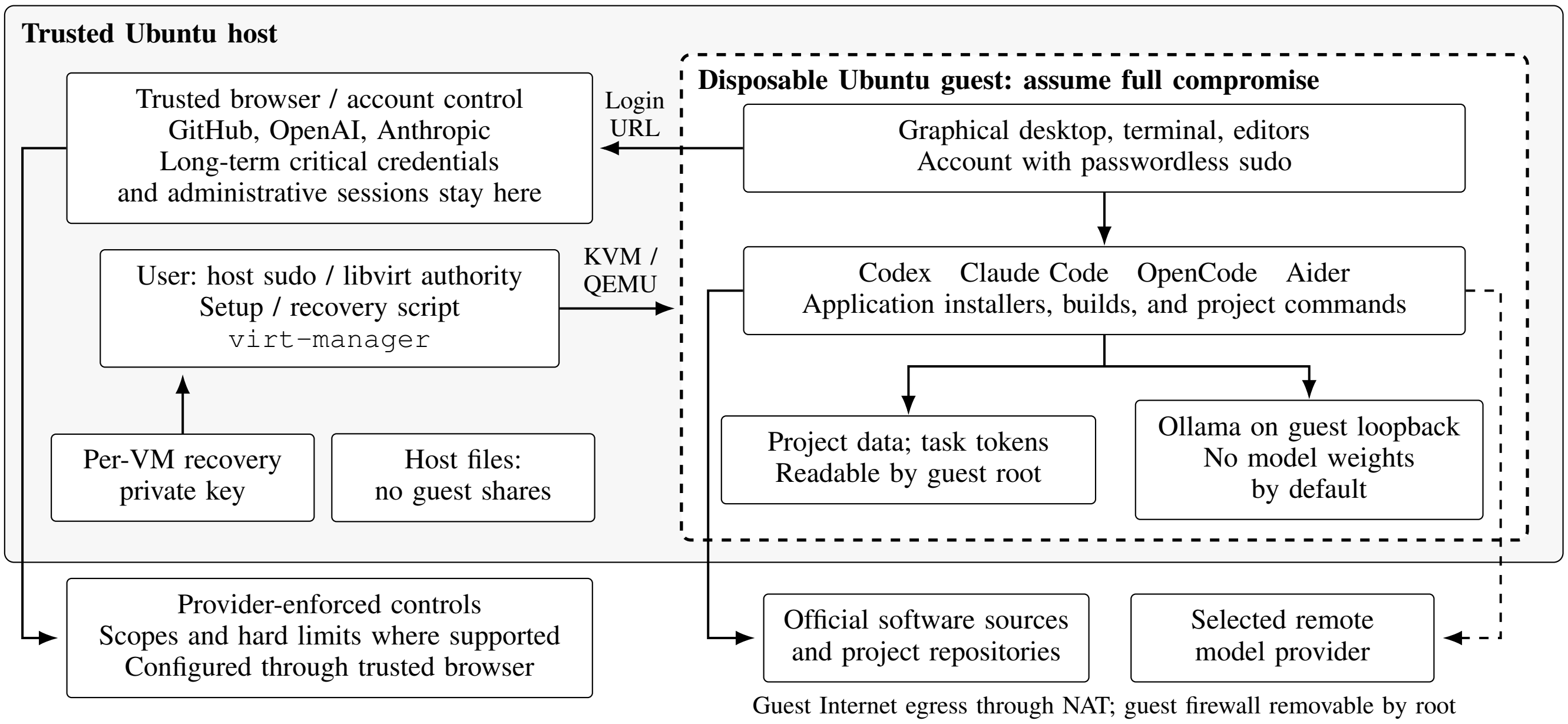
\begin{figure*}[t]
\centering
\begin{tikzpicture}[x=1cm,y=1cm,font=\small,
  box/.style={draw,rounded corners=1.5pt,align=center,inner sep=5pt,fill=white},
  flow/.style={-{Latex[length=2.2mm]},line width=.75pt}]
\draw[rounded corners=3pt,fill=black!3] (0,-.5) rectangle (17.5,5.75);
\node[anchor=north west,inner sep=0pt,font=\bfseries\small] at (.18,5.56) {Trusted Ubuntu host};
\draw[rounded corners=3pt,line width=1pt,dashed,fill=white] (7.6,-.25) rectangle (17.1,5.2);
\node[anchor=north west,inner sep=0pt,font=\bfseries\small] at (7.78,5.04) {Disposable Ubuntu guest: assume full compromise};
\node[box,text width=5.55cm] (browser) at (3.65,4.15) {Trusted browser / account control\\GitHub, OpenAI, Anthropic\\Long-term critical credentials\\and administrative sessions stay here};
\node[box,text width=4.8cm] (control) at (3.65,2.35) {User: host sudo / libvirt authority\\Setup / recovery script\\\texttt{virt-manager}};
\node[box,text width=2.6cm] (recovery) at (2.0,.45) {Per-VM recovery\\private key};
\node[box,text width=2.6cm] (hostfiles) at (5.15,.45) {Host files:\\no guest shares};
\draw[flow,shorten >=2pt] (recovery.north) -- (recovery.north |- control.south);
\node[box,text width=7.75cm] (desktop) at (12.35,4.15) {Graphical desktop, terminal, editors\\Account with passwordless sudo};
\node[box,text width=7.75cm] (agents) at (12.35,2.55) {Codex\quad Claude Code\quad OpenCode\quad Aider\\Application installers, builds, and project commands};
\node[box,text width=3.85cm] (data) at (10.15,.65) {Project data; task tokens\\Readable by guest root};
\node[box,text width=3.55cm] (model) at (14.65,.65) {Ollama on guest loopback\\No model weights by default};
\draw[flow] (desktop) -- (agents);
\coordinate (branch) at (12.35,1.7);
\draw[line width=.75pt] (agents.south) -- (branch);
\draw[flow] (branch) -| (data.north);
\draw[flow] (branch) -| (model.north);
\draw[flow,shorten >=2pt] (control.east) -- node[above,align=center,font=\footnotesize] {KVM /\\QEMU} (7.6,2.35);
\draw[flow,shorten >=2pt] (desktop.west) -- node[pos=.72,above,align=center,font=\footnotesize] {Login\\URL} (browser.east);
\node[box,text width=5.55cm] (policy) at (3.65,-1.35) {Provider-enforced controls\\Scopes and hard limits where supported\\Configured through trusted browser};
\draw[flow,shorten >=2.5pt] (browser.west) -- (.2,4.15) -- (.2,-1.35) -- (policy.west);
\node[box,text width=3.3cm] (sources) at (10.35,-1.35) {Official software sources\\and project repositories};
\node[box,text width=3.05cm] (provider) at (14.35,-1.35) {Selected remote\\model provider};
\draw[flow,shorten >=2.5pt] (agents.west) -- (7.9,2.55) -- (7.9,-1.35) -- (sources.west);
\draw[flow,dashed,shorten >=2.5pt] (agents.east) -- (16.8,2.55) -- (16.8,-1.35) -- (provider.east);
\node[align=center,font=\footnotesize,text width=9.0cm] at (12.35,-2.13) {Guest Internet egress through NAT; guest firewall removable by root};
\end{tikzpicture}
\caption{Architecture and credential placement. Recovery access is separate from provider-account control. Guest applications receive usable tokens; provider limits require separate configuration and credentials unable to change them. The login-URL arrow shows guest-to-host transfer through the shared clipboard. The dashed guest boundary denotes assumed compromise; dashed model access is optional.}
\label{fig:architecture}
\end{figure*}

Our ready-to-use reference implementation runs on x86-64 Ubuntu 24.04 or 26.04 hosts and provisions Ubuntu 26.04 guests.
Fig.~\ref{fig:architecture} illustrates the architecture,
while Fig.~\ref{fig:managerworker} shows a scenario involving manager-worker deployment.
Source code and setup instructions are available at
\url{https://github.com/yutakang/agent-vm}~\cite{kvmagent_github}.

\subsection{Autonomy versus containment}\label{sec:imp-autonomy}
\subsubsection{Threat assumptions and risk coverage}
For the autonomy goal (Section~\ref{sec:goal-autonomy}), the VM boundary separates guest authority from host OS resources and unrelated files and credentials.
We assume arbitrary guest commands, including as root; guest projects, sessions, and supplied credentials are exposed.
The user, physical machine, firmware, host kernel, KVM, QEMU, libvirt, and administration tools form the trusted foundation.

OpenAI and Anthropic provide OS-enforced sandboxes whose coverage depends on the application, configuration, and integration paths~\cite{codexapprovals,claudepermissions,claudesandbox}.
For host containment, we rely on KVM/QEMU rather than these application-specific sandboxes, keeping the isolation mechanism independent of the chosen coding-agent application (Section~\ref{sec:goal-sovereignty}).

We retain Internet access by default so that agents can obtain software, dependencies, and data.
By default, the guest firewall restricts access to local devices and services by blocking private and link-local IP address ranges.
An agent with root access inside the guest can change or disable these rules.
Safeguards supplied by model providers and developers of coding-agent applications can provide complementary checks on agent actions.
Network restrictions intended to survive full guest compromise require enforcement outside the guest, with policies adapted to the deployment's network topology and the services agents need to access.
Users can configure host-enforced egress policies or disable guest networking from the host when the task permits.

Table~\ref{tab:risks} summarizes threat scenarios, corresponding controls and user practices, and residual exposure.
While virtualization exploits and side channels remain residual risks, we omit host-directory shares, host container sockets, and physical-device passthrough to limit paths to host authority~\cite{nistvirtualization}.

\begin{table*}[t]
\caption{Risk coverage under KVM-Agent's stated trust assumptions.}
\label{tab:risks}
\centering
\small
\renewcommand{\arraystretch}{1.12}
\begin{tabularx}{\textwidth}{@{}>{\raggedright\arraybackslash}p{3.1cm}YY@{}}
\toprule
Scenario & Controls and user practices & Residual exposure \\
\midrule
Destroy or corrupt important data & No host-home/project mount; normally one project per disposable guest. & All co-located projects remain exposed. Backups and version control are separate requirements. \\
Steal data or credentials & Primary account credentials, admin sessions, and recovery/controller keys stay outside the guest; SSH-agent forwarding is disabled. & Guest tokens can be stolen and used within their granted scope. Internet access permits exfiltration; shared clipboard contents can cross the boundary. \\
Attack other machines & Optional, user-configured directional Tailscale access rules; guest firewall defaults restrict private ranges. & Guest root can remove its firewall. NAT is not an egress allowlist; Internet and permitted peers remain reachable. \\
Gain host authority & KVM/QEMU boundary; no host socket or device passthrough. & Virtualization defects and compromised host administration can invalidate separation. \\
Exhaust resources or incur costs & VM RAM, vCPU, and disk capacity; separately configured provider hard limits where supported. & Guest credentials must lack limit-changing authority. Alerts are not caps; no comprehensive I/O/bandwidth limits; thin disks can exhaust host storage. \\
Spoof a peer or export malicious content & SSH key pinning; console fingerprint check for remote access; host-side quarantine of retrieved files. & Initial recovery trusts the first SSH host key. Clipboard, terminal output, and exported code still need scrutiny. \\
Erase or falsify activity records & Provisioning and job logs aid diagnosis. & Guest-root-writable records are not a tamper-evident audit trail. \\
\bottomrule
\end{tabularx}

\end{table*}

\subsubsection{Interaction and external authority}
As discussed, we assume that guest VMs may be fully compromised; however, agents still need access to online services, including paid services.
We therefore separate guest service access from account administration and recovery.
GitHub, OpenAI, and Anthropic account passwords, passkeys, recovery secrets, and administrative browser sessions stay on the trusted host or remote controller (Fig.~\ref{fig:remote}).
Where supported, cross-device authorization grants guest service access while users sign in on a trusted device; Codex requires enabling device-code authentication~\cite{ghauth,codexauth,claudeauth}.
The guest nevertheless receives access/session tokens, possibly refresh tokens, or task API keys; an attacker with root access inside the guest can steal and use them.
Where supported, guests should receive narrowly scoped, revocable task credentials with short lifetimes.

For OpenAI and Anthropic services, financial exposure can be reduced through subscription-based access without paid overages, or API access subject to provider-enforced spending caps where available.
Budget alerts alone do not stop usage, and cap enforcement varies by product and plan~\cite{openaispend,anthropiclimits,githubbudgets}.
Users administer billing, permissions, and usage limits from the trusted device; guest credentials must not authorize changes to these settings.

Under the one-project-per-VM convention, a GitHub fine-grained token can be scoped to the project's repositories and required permissions, with an expiry~\cite{githubtokens}.
Branch protection and human review provide additional controls.

Clipboard sharing supports a common workflow for transferring provider login URLs to the trusted browser, but can also expose host clipboard contents to the guest.
In our experience, this convenience justifies retaining clipboard sharing.

\subsection{Provider reliance versus user sovereignty}\label{sec:imp-sovereignty}

For the sovereignty goal (Section~\ref{sec:goal-sovereignty}), provisioning the execution environment independently of the coding-agent application preserves application and provider choice.
By default, setup installs Codex, Claude Code, OpenCode, Aider, GitHub CLI, and Ollama.
Some services reserve particular features for paid plans, but preinstalled alternatives make it easier to switch applications or providers and reduce dependence on any single vendor.

For users with sufficient computing resources, Ollama enables model inference inside the guest, keeping inference inputs local and avoiding reliance on external LLM services.
The reference implementation supports CPU inference without GPU passthrough.
Using a hosted model instead shifts computation elsewhere and sends the configured project context to that service.

\subsection{Functionality versus operational cost}\label{sec:imp-operation}
To address the operational goal (Section~\ref{sec:goal-operation}), \tool{} automates most setup and provisioning steps and provides tools for recovery, remote access, and running jobs on worker VMs.
Manual input remains necessary for user-specific choices, credentials, and trust decisions.

\subsubsection{Provisioning and recovery}
Setup verifies the Ubuntu image against Ubuntu's signed checksums, creates the VM with a temporary disk containing its initial configuration~\cite{nocloud}, and installs the desktop and development tools.
The host checks that provisioning and the required reboot have completed, removes cached setup data, and confirms that the configuration disk is disconnected before deleting it.
Version records aid diagnosis, but updates to upstream installers prevent byte-for-byte reproducibility.

Recovery access uses a dedicated SSH key for each VM, with the private key kept on the host.
Users should review and export needed results before replacing a VM.
After suspected compromise, users should use the trusted host or controller to revoke or rotate service credentials used inside the guest; deleting a VM does not revoke stolen tokens.

\subsubsection{Remote access and persistent sessions}\label{sec:remote}
The basic setup runs on a single computer; remote access is optional.
For users with multiple computers, a laptop can serve as the controller for a VM running on a more powerful desktop (Fig.~\ref{fig:remote}).
The guest includes \code{tmux} to keep terminal sessions running when a remote connection is interrupted, allowing users to reconnect to ongoing work~\cite{tmux}.
These sessions end if the guest reboots.

OpenSSH provides authenticated remote terminal access, while Tailscale connects the controller and guests through an encrypted private network~\cite{tailscaleconnections}.
The physical VM hosts remain outside this private network.
As with the online services discussed above, users authorize a guest by opening its Tailscale sign-in URL in a browser on the trusted host or controller, keeping account credentials and administrative browser sessions outside the guest.
Tailscale simplifies connection setup but introduces reliance on its managed coordination service.
Self-managed WireGuard is an alternative requiring more setup effort~\cite{wireguard}.

Users authorize devices and configure access rules that allow the controller to connect to guests while blocking guest-initiated connections to the controller and unrelated devices~\cite{tailscaleacls}.
Users must replace the default allow-all policy with rules enforced by receiving devices independently of guest settings; setup does not configure this policy.
The rules govern the private network; access through other networks requires separate controls.

\tool{} includes a setup script for using a macOS computer as a trusted remote controller.
We have tested this configuration.

\subsubsection{Manager--worker execution}\label{sec:managerworker}

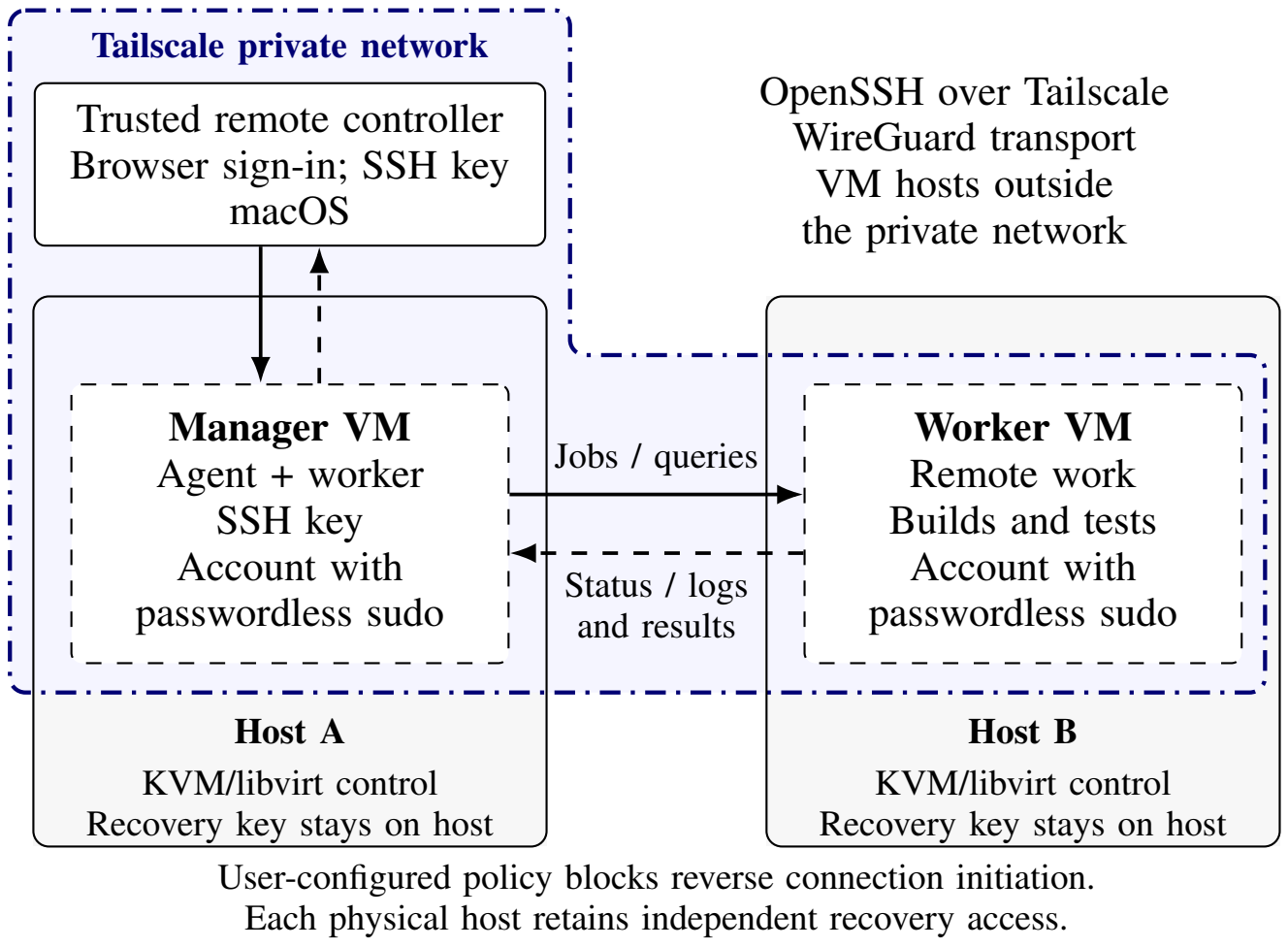
\begin{figure}[t]
\centering
\begin{tikzpicture}[x=1cm,y=1cm,font=\footnotesize,
  box/.style={draw,rounded corners=2pt,align=center,
    inner sep=4pt,fill=white},
  flow/.style={-{Latex[length=1.8mm]},line width=.7pt}]

\path[fill=black!3] (0,-.9) rectangle (3.5,-4.65);
\path[fill=black!3] (5,-.9) rectangle (8.5,-4.65);

\draw[rounded corners=3pt,draw=blue!45!black,line width=.8pt,
  dash pattern=on 4pt off 2pt on .8pt off 2pt,fill=blue!4]
  (-.16,1.05) -- (3.66,1.05) -- (3.66,-1.30)
  -- (8.40,-1.30) -- (8.40,-3.60)
  -- (-.16,-3.60) -- cycle;
\node[font=\bfseries\scriptsize,text=blue!45!black]
  at (1.75,.79) {Tailscale private network};

\draw[rounded corners=3pt] (0,-.9) rectangle (3.5,-4.65);
\draw[rounded corners=3pt] (5,-.9) rectangle (8.5,-4.65);

\node[box,text width=3.2cm] (controller) at (1.75,0)
  {Trusted remote controller\\Browser sign-in; SSH key\\macOS};
\node[align=center,text width=3.6cm] at (6.35,0)
  {OpenSSH over Tailscale\\WireGuard transport\\
   VM hosts outside the private network};

\node[box,dashed,text width=2.7cm,minimum height=1.9cm]
  (manager) at (1.75,-2.45)
  {\textbf{Manager VM}\\Agent + worker\\SSH key\\
   Account with\\passwordless sudo};
\node[box,dashed,text width=2.7cm,minimum height=1.9cm]
  (worker) at (6.75,-2.45)
  {\textbf{Worker VM}\\Remote work\\Builds and tests\\
   Account with\\passwordless sudo};

\draw[flow] ([xshift=-2mm]controller.south)
  -- ([xshift=-2mm]manager.north);
\draw[flow,dashed] ([xshift=2mm]manager.north)
  -- ([xshift=2mm]controller.south);
\draw[flow] ([yshift=2mm]manager.east)
  -- node[above,font=\scriptsize]{Jobs / queries}
  ([yshift=2mm]worker.west);
\draw[flow,dashed] ([yshift=-2mm]worker.west)
  -- node[below,font=\scriptsize,align=center]{Status / logs\\and results}
  ([yshift=-2mm]manager.east);

\node[align=center,font=\scriptsize,anchor=north] at (1.75,-3.65)
  {\textbf{Host A}\\[2pt]KVM/libvirt control\\Recovery key stays on host};
\node[align=center,font=\scriptsize,anchor=north] at (6.75,-3.65)
  {\textbf{Host B}\\[2pt]KVM/libvirt control\\Recovery key stays on host};

\node[align=center,text width=8.3cm,font=\scriptsize] at (4.25,-5.02)
  {User-configured policy blocks reverse connection initiation.\\
   Each physical host retains independent recovery access.};
\end{tikzpicture}
\caption{Remote control and delegation. Solid arrows show SSH initiation; dashed arrows show replies. The controller and guests join the private network established using Tailscale; physical VM hosts do not. Provider-account administration and controller keys stay on the controller; recovery keys stay on hosts. The manager holds worker-access keys.}
\label{fig:remote}
\label{fig:managerworker}
\end{figure}

Users with spare computers may want to offload long-running builds or tests while continuing development in their main VM.
In this optional arrangement, the main VM acts as the manager and submits jobs over OpenSSH to worker VMs on other machines (Fig.~\ref{fig:managerworker}).
In our deployment, we configure the manager to access 
accounts inside worker VMs with passwordless sudo for software installation, dependency repair, and job execution as guest root.
Users retain administration of the physical hosts and keep recovery keys on those hosts.

For build and test jobs, model-service tokens can remain on the manager, while primary account credentials stay on the trusted host or controller.
Access rules for the private network should allow the manager to connect to workers and receive replies, while blocking workers from initiating connections to the manager, controller, or unrelated devices.
Because a compromised manager can also control its authorized workers, users should dedicate the manager and its workers to the same project.

To accelerate development of the next version of \tool{}, we use this manager--worker setup to run independent development and testing tasks in parallel across three VMs hosted on older laboratory computers.

\section{Related Work}\label{sec:related}
SWE-agent studies agent--computer interfaces~\cite{yang2024swe}; OpenHands integrates agent development and sandboxed execution~\cite{wang2026openhands}.
Firecracker targets serverless workloads~\cite{agache2020firecracker}, while Docker Sandboxes provides coding agents with sudo inside microVMs and host-side network and credential controls~\cite{dockersandboxes}.
Anthropic also describes VM-based code execution for Cowork~\cite{anthropiccontainment}.
\tool{} builds on existing virtualization to automate a user-managed graphical Linux workflow with a choice of coding-agent applications, model services, and optional remote execution.

The Nitro Isolation Engine provides formal assurance for specified VM-management mechanisms~\cite{nitroverified}; seL4 has machine-checked kernel proofs under explicit assumptions~\cite{klein2014sel4,sel4assumptions}.
Agentic-seL4 proposes an agent sandbox with dynamic permission and information-flow controls~\cite{agenticsel4}; its planned guarantees differ from established kernel results.
This work on stronger isolation assurance complements our focus on deployment, provider choice, and operational cost.

A potential concern is the performance overhead introduced by KVM/QEMU in \tool{}.
Independent studies report KVM performance at 95.9\% of bare metal for Linux kernel compilation~\cite{giallorenzo2021costs} and 6\% overhead for SPEC ACCEL 1.2 OpenMP workloads~\cite{tharwani2026overhead}.
Although these studies do not evaluate \tool{} itself, their results suggest that VM isolation can impose modest execution overhead for such workloads.

\section{Discussion and Conclusion}\label{sec:conclusion}

This paper examined three trade-offs in deploying coding agents: \emph{agent autonomy versus containment}, \emph{provider reliance versus user sovereignty}, and \emph{functionality versus operational cost}.
We presented \tool{} as an open-source reference implementation of the resulting design.
Agents administer replaceable project VMs, while users retain host administration, recovery, and account control.
The implementation preserves choices among coding-agent applications and model services, supports familiar development tools and optional remote execution, and automates most setup and provisioning steps.

This balance still requires VM resources, software maintenance, and user decisions about credentials and network access. 
In practice, however, we have found these demands manageable for our small research team. 
We have used KVM-Agent in four projects: ProofStudio~\cite{proofstudio}, 
the neural adaptation of AbductionProver~\cite{abduction_prover}, 
a cross-prover proof-automation tool, 
and the next version of KVM-Agent itself. 
In these projects, we use commercial LLM services while restricting agents’ authority over our machines to project-specific VMs. 
We retain the option to switch to models hosted on Czech academic infrastructure (e-INFRA CZ~\cite{einfra_llm}), reducing our dependence on commercial providers.

Future work will evaluate setup and recovery effort, execution performance, and the effort required to switch coding-agent applications or model providers.
It will also assess host and network controls under full guest compromise.
These evaluations will help refine the choices embodied in \tool{}, which provides a largely automated way to deploy autonomous coding agents while retaining user control under resource constraints.

\section*{Acknowledgment}
ChatGPT, Codex, Claude Code, OpenCode, and Kimi-K3 assisted with brainstorming, manuscript drafting and revision, literature survey, and software implementation.

\bibliographystyle{IEEEtran}
\bibliography{references}

@inproceedings{yang2024swe,
 author={John Yang and Carlos E. Jimenez and Alexander Wettig and Kilian Lieret and Shunyu Yao and Karthik Narasimhan and Ofir Press},
 title={{SWE}-agent: Agent-Computer Interfaces Enable Automated Software Engineering},
 booktitle={Advances in Neural Information Processing Systems},
 volume={37}, year={2024},
 url={https://proceedings.neurips.cc/paper_files/paper/2024/hash/5a7c947568c1b1328ccc5230172e1e7c-Abstract-Conference.html}
}

@inproceedings{greshake2023indirect,
  author = {Greshake, Kai and Abdelnabi, Sahar and
            Mishra, Shailesh and Endres, Christoph and
            Holz, Thorsten and Fritz, Mario},
  title = {Not What You've Signed Up For:
           Compromising Real-World {LLM}-Integrated
           Applications with Indirect Prompt Injection},
  booktitle = {Proceedings of the 16th ACM Workshop
               on Artificial Intelligence and Security},
  year = {2023},
  pages = {79--90},
  publisher = {ACM},
  doi = {10.1145/3605764.3623985}
}

@inproceedings{kivity2007kvm,
 author={Avi Kivity and Yaniv Kamay and Dor Laor and Uri Lublin and Anthony Liguori},
 title={{kvm}: The {Linux} Virtual Machine Monitor},
 booktitle={Proceedings of the Linux Symposium}, volume={1},
 year={2007}, pages={225--230},
 url={https://www.kernel.org/doc/ols/2007/ols2007v1-pages-225-230.pdf}
}

@inproceedings{bellard2005qemu,
 author={Fabrice Bellard},
 title={{QEMU}, a Fast and Portable Dynamic Translator},
 booktitle={USENIX Annual Technical Conference, FREENIX Track},
 year={2005}, pages={41--46}, publisher={USENIX Association}
}

@misc{libvirtoverview,
 author={{libvirt project}}, title={The virtualization {API}},
 howpublished={Online documentation},
 url={https://libvirt.org/},
 note={Accessed October 31, 2026}
}

@misc{nocloud,
 author={{Canonical}}, title={{NoCloud} datasource},
 howpublished={cloud-init documentation},
 url={https://docs.cloud-init.io/en/latest/reference/datasources/nocloud.html},
 note={Accessed October 26, 2026}
}

@inproceedings{wang2026openhands,
 author={Xingyao Wang and Simon Rosenberg and Juan Michelini and Calvin Smith and Hoang H. Tran and Engel Nyst and Rohit Malhotra and Xuhui Zhou and Valerie Chen and Robert Brennan and Graham Neubig},
 title={The {OpenHands} Software Agent {SDK}: A Composable and Extensible Foundation for Production Agents},
 booktitle={Proceedings of Machine Learning and Systems}, volume={8}, year={2026},
 url={https://proceedings.mlsys.org/paper_files/paper/2026/hash/8ae9cf363ea625161f885b798c1f1f78-Abstract-Conference.html}
}

@inproceedings{agache2020firecracker,
 author={Alexandru Agache and Marc Brooker and Andreea Florescu and Alexandra Iordache and Anthony Liguori and Rolf Neugebauer and Phil Piwonka and Diana-Maria Popa},
 title={Firecracker: Lightweight Virtualization for Serverless Applications},
 booktitle={17th USENIX Symposium on Networked Systems Design and Implementation (NSDI)},
 year={2020}, pages={419--434}, publisher={USENIX Association},
 url={https://www.usenix.org/conference/nsdi20/presentation/agache}
}

@misc{dockersandboxes,
 author={{Docker}}, title={{Docker Sandboxes}: Isolation layers},
 howpublished={Online documentation},
 url={https://docs.docker.com/ai/sandboxes/security/isolation/},
 note={Accessed October 26, 2026}
}

@misc{linuxcredentials,
 author={{Linux man-pages project}}, title={credentials(7)},
 howpublished={Linux manual pages},
 url={https://man7.org/linux/man-pages/man7/credentials.7.html},
 note={Accessed October 26, 2026}
}

@techreport{nistcontainers,
 author={Murugiah Souppaya and John Morello and Karen Scarfone},
 title={Application Container Security Guide},
 institution={National Institute of Standards and Technology},
 number={SP 800-190}, year={2017},
 doi={10.6028/NIST.SP.800-190},
 url={https://doi.org/10.6028/NIST.SP.800-190}
}

@misc{dockersecurity,
 author={{Docker}}, title={{Docker Engine} security},
 howpublished={Online documentation},
 url={https://docs.docker.com/engine/security/},
 note={Accessed October 26, 2026}
}

@misc{dockerrootless,
 author={{Docker}}, title={Rootless mode},
 howpublished={Online documentation},
 url={https://docs.docker.com/engine/security/rootless/},
 note={Accessed October 26, 2026}
}

@misc{podman,
 author={{Podman project}}, title={podman(1)},
 howpublished={Podman documentation},
 url={https://docs.podman.io/en/latest/markdown/podman.1.html},
 note={Accessed October 26, 2026}
}

@misc{tailscaleconnections,
 author={{Tailscale}}, title={Connection types},
 howpublished={Online documentation},
 url={https://tailscale.com/docs/reference/connection-types},
 note={Accessed October 26, 2026}
}

@misc{tailscaleacls,
 author={{Tailscale}}, title={Manage permissions using {ACLs}},
 howpublished={Online documentation},
 url={https://tailscale.com/docs/features/access-control/acls},
 note={Accessed October 26, 2026}
}

@misc{wireguard,
 author={Jason A. Donenfeld}, title={{WireGuard}: Next Generation Kernel Network Tunnel},
 howpublished={Technical whitepaper, draft revision e2da747},
 year={2020}, month={June~1,},
 url={https://www.wireguard.com/papers/wireguard.pdf},
 note={Accessed October 26, 2026}
}

@techreport{nitroverified,
 author={{Amazon Web Services}}, title={{Nitro Isolation Engine} Whitepaper},
 institution={Amazon Web Services}, year={2026}, month=jun,
 url={https://d1.awsstatic.com/onedam/marketing-channels/website/aws/en_US/whitepapers/compliance/nitro-isolation-engine-whitepaper.pdf}
}

@article{klein2014sel4,
 author={Gerwin Klein and June Andronick and Kevin Elphinstone and Toby Murray and Thomas Sewell and Rafal Kolanski and Gernot Heiser},
 title={Comprehensive Formal Verification of an {OS} Microkernel},
 journal={ACM Transactions on Computer Systems}, volume={32}, number={1},
 pages={2:1--2:70}, year={2014}, doi={10.1145/2560537}
}

@misc{sel4assumptions,
 author={{seL4 project}}, title={What the Proofs Assume},
 url={https://sel4.systems/Verification/assumptions.html},
 note={Accessed October 26, 2026}
}

@misc{agenticsel4,
 author={{University of Surrey}},
 title={Surrey leads {\pounds}3.2m project in collaboration with {Sheffield} and {Melbourne} to develop provably secure sandbox for {AI} agents},
 howpublished={Project announcement}, year={2026}, month=oct,
 url={https://www.surrey.ac.uk/news/surrey-leads-ps32m-project-collaboration-sheffield-and-melbourne-develop-provably-secure-sandbox-ai}
}

@article{giallorenzo2021costs,
 author={Saverio Giallorenzo and Jacopo Mauro and Martin Gyde Poulsen and Filip Siroky},
 title={Virtualization Costs: Benchmarking Containers and Virtual Machines Against Bare-Metal},
 journal={SN Computer Science}, volume={2}, number={5}, year={2021},
 note={Art. no. 404},
 doi={10.1007/s42979-021-00781-8}
}

@inproceedings{tharwani2026overhead,
 author={Jay Tharwani and Shobhit Aggarwal and Arnab A. Purkayastha},
 title={Quantifying Virtualization Overhead for {CPU}-Intensive {HPC} Workloads in Hypervisors: A Comparative Study of {KVM}, {Proxmox VE}, {Xen}, and {Hyper-V}},
 booktitle={IEEE SoutheastCon}, pages={1--6}, year={2026},
 doi={10.1109/SoutheastCon63549.2026.11476478}
}

@techreport{nistvirtualization,
  author = {Karen Scarfone and Murugiah Souppaya and Paul Hoffman},
  title = {Guide to Security for Full Virtualization Technologies},
  institution = {National Institute of Standards and Technology},
  number = {SP 800-125},
  year = {2011},
  doi = {10.6028/NIST.SP.800-125}
}

@misc{tmux,
  author = {{tmux contributors}},
  title = {{tmux}: Getting Started},
  howpublished = {Project documentation},
  url = {https://github.com/tmux/tmux/wiki/Getting-Started},
  note = {Accessed October 27, 2026}
}

@misc{codexapprovals,
  author={{OpenAI}}, title={Agent approvals \& security},
  howpublished={Codex documentation},
  url={https://developers.openai.com/codex/agent-approvals-security},
  note={Accessed October 27, 2026}
}

@misc{claudepermissions,
  author={{Anthropic}}, title={Choose a permission mode},
  howpublished={Claude Code documentation},
  url={https://code.claude.com/docs/en/permission-modes},
  note={Accessed October 27, 2026}
}

@misc{claudesandbox,
  author={{Anthropic}}, title={Configure the sandboxed {Bash} tool},
  howpublished={Claude Code documentation},
  url={https://code.claude.com/docs/en/sandboxing},
  note={Accessed October 27, 2026}
}

@misc{anthropiccontainment,
  author={{Anthropic}}, title={How we contain {Claude} across products},
  howpublished={Engineering report}, year={2026}, month=may,
  url={https://www.anthropic.com/research/how-we-contain-claude},
  note={Accessed October 27, 2026}
}

@misc{ghauth,
  author={{GitHub}}, title={{gh auth login}},
  howpublished={GitHub CLI manual},
  url={https://cli.github.com/manual/gh_auth_login},
  note={Accessed October 29, 2026}
}

@misc{codexauth,
  author={{OpenAI}}, title={Authentication},
  howpublished={Codex documentation},
  url={https://developers.openai.com/codex/auth},
  note={Accessed October 29, 2026}
}

@misc{claudeauth,
  author={{Anthropic}}, title={Authentication},
  howpublished={Claude Code documentation},
  url={https://code.claude.com/docs/en/authentication},
  note={Accessed October 29, 2026}
}

@misc{openaispend,
  author={{OpenAI}}, title={Spend limits},
  howpublished={API documentation},
  url={https://developers.openai.com/api/docs/guides/spend-limits},
  note={Accessed October 29, 2026}
}

@misc{anthropiclimits,
  author={{Anthropic}}, title={Rate limits},
  howpublished={Claude Platform documentation},
  url={https://platform.claude.com/docs/en/api/rate-limits},
  note={Accessed October 29, 2026}
}

@misc{githubbudgets,
  author={{GitHub}}, title={Budgets and alerts},
  howpublished={Billing documentation},
  url={https://docs.github.com/en/billing/concepts/budgets-and-alerts},
  note={Accessed October 29, 2026}
}

@misc{githubtokens,
  author={{GitHub}}, title={Managing your personal access tokens},
  howpublished={GitHub documentation},
  url={https://docs.github.com/en/authentication/keeping-your-account-and-data-secure/managing-your-personal-access-tokens},
  note={Accessed October 29, 2026}
}

@misc{proofstudio,
  author       = {Yutaka Nagashima},
  title        = {{ProofStudio}: Five Proof Assistants in the Browser},
  howpublished = {arXiv preprint arXiv:2611.05540},
  year         = {2026},
  month        = nov,
  doi          = {10.48550/arXiv.2611.05540},
  url          = {https://arxiv.org/abs/2611.05540}
}

@article{abduction_prover,
  author       = {Yutaka Nagashima and
                  Daniel Sebastian Goc},
  title        = {Abduction Prover in {I}sabelle/{HOL}},
  journal      = {CoRR},
  volume       = {abs/2606.04877},
  year         = {2026},
  url          = {https://doi.org/10.48550/arXiv.2606.04877},
  doi          = {10.48550/ARXIV.2606.04877},
  eprinttype   = {arXiv},
  eprint       = {2606.04877},
  bibsource    = {dblp computer science bibliography, https://dblp.org}
}

@misc{einfra_llm,
  author       = {{CERIT-SC}},
  title        = {{AIaaS} Introduction},
  howpublished = {{e-INFRA CZ} documentation},
  url          = {https://docs.cerit.io/en/docs/ai-as-a-service/introduction},
  note         = {Accessed: 2026-11-09}
}

@misc{kvmagent_github,
  author       = {Yutaka Nagashima},
  title        = {{KVM-Agent}: A Graphical {Ubuntu} {VM} for Coding Agents},
  howpublished = {GitHub repository},
  year         = {2026},
  url          = {https://github.com/yutakang/agent-vm},
  note         = {Accessed: 2026-11-09}
}
\end{document}